# Gesture analysis for physics education researchers

Rachel E. Scherr
Physics Education Research Group, Department of Physics
University of Maryland, College Park

ABSTRACT

Systematic observations of student gestures can not only fill in gaps in students' verbal expressions, but can also offer valuable information about student ideas, including their source, their novelty to the speaker, and their construction in real time. This paper provides a review of the research in gesture analysis that is most relevant to physics education researchers and illustrates gesture analysis for the purpose of better understanding student thinking about physics.

## I.  INTRODUCTION

A student trying to decide whether an object in projectile motion has zero velocity at the top of its flight says "Wouldn't it just fall straight down then if it was like –Wmp! Psh." The transcript of her verbalizations, in this case, is insufficient for us to understand her idea; in order to understand what the student is "saying," we need to see her gesture. Gestures are one channel of the rich stream of data that is the basis of our investigations of student thinking. In some cases, gestures are clearly indispensable for filling in verbal "blanks" in speech. However, systematic observations of gestures can also offer other valuable information about student ideas, including their source, their novelty to the speaker, and their construction in real time. Research on gesture analysis appears in literature from cognitive science, linguistics, and learning sciences, and weaves together insights from these diverse traditions.

The purpose of this paper is to show how physics education researchers may analyze gestures for the purpose of better understanding student thinking. Physics educators may benefit from attending to gestures as well, and this paper draws attention to aspects of gesture analysis that are particularly relevant for instructors. In what follows I use two different episodes of student conversation as contexts for introducing gesture research. In the first episode (Section III), one key gesture is necessary for understanding the content of a particular student's idea, and also gives us evidence of the source of that idea – that it's something she constructed herself, rather than something conveyed to her by a teaching assistant. This first episode then serves as a context for exploring literature that uses gestures as indicators of the novelty of student ideas as well as their construction in real time (Section IV). A second episode from a different physics context then serves to illustrate two more detailed analyses (Section V): one in which a series of gestures provides evidence of the decreasing novelty of student ideas, and another in which gestures may provide students with particular cognitive resources for analyzing a collision.

## II.  BACKGROUND OF GESTURE RESEARCH

Much gesture research shares common definitions and foundational assumptions. In what follows I review the basic ideas on which the majority of the literature rests, including what actions are identified as gestures and the role of gestures in communication and thinking.

A. Identifying gestures

Gestures are the spontaneous hand movements of individual speakers – movements that are directly tied to speech and are created at the moment of speaking.[1-4] Such movements are in the service of communication and are in that sense deliberate; however, speakers are often not aware of having made them, and in that sense they are unwitting. Various classification systems divide gestures into those in which the hands represent objects that are in the scene that speech presents (*e.g.,* the projectile in the situation described above), those in which the hands represent abstract concepts by taking a concrete form (*e.g.,* a narrative genre represented by an enclosure), and nonrepresentational hand motions (*e.g.,* pointing).[5] Actions such as habitual hand movements (rubbing one's chin, or smoothing one's hair) and functional acts on objects (opening a jar, or writing) lack communicative intent and are therefore not gestures.[6] Gestures also do not include culturally agreed upon signs such as "thumbs up," which are independent of speech, and sign languages, which are constrained by grammatical and phonological systems and are disrupted by speech.[7]

B. Gesture's role in communication and in thinking

Gestures "participate in communication, yet are not part of a codified system. As such, they are free to take on forms that speech cannot assume and are consequently free to reveal meanings that speech cannot accommodate."[8] Gestures are not limited by the linguistic and phonological constraints of spoken or signed languages. Instead, they often provide a unique view of speakers' mental representations of objects, phenomena, ideas, and so on. The fact that gestures are typically performed unconsciously makes them especially tempting as a potential window into speakers' unedited thinking. For example, a person may say, "I have a question," while forming one hand into the shape of a cup. The cup may represent the question as a bounded, supportable object, or may imply that the answer is a substance that could be placed in the hand.[9] In either case, the imagery is metaphoric and not part of the spoken representation. Later appearances of cup-shaped hands might refer to entirely different entities, but might also retain some of their initial meaning, providing imagistic resonances that might influence later expressions. (Examples of this effect appear in Section V below.) In another example, a student described in Section V below says, "if we actually saw [the collision]," while moving his pointed index finger outward from his eye. Such a gesture reveals a mental image of "seeing" as involving an emission from the eye, an image that the speaker would not be likely to represent verbally but which apparently persists at some level.[10]

Gesture, language, and thought occur simultaneously, sharing both meaning and context, yet appearing in different media. They are not separable from one another, but are different aspects of a single process that simultaneously involves physical action, cognition, and neurological events.[11] Gestures accompany speech and reveal thought, in the sense that they communicate images, actions, and so on that may or may not be expressed verbally.[12] Gestures are also active participants in speaking and thinking: they fuel speech and thought by providing shared meaning in a mode unlike speech. Specific examples of this integration and the mechanisms by which gesture, speech, and thought interact are provided in Section IV.[13]

## III. TRAJECTORY EPISODE: GESTURES AS EVIDENCE OF THE CONTENT AND SOURCE OF STUDENT IDEAS

As suggested in the introduction, gestures may express aspects of meaning that are not expressed in speech. Below, I present an episode in which a student's gesture is the main source of information about not only the content, but also the source of her idea.[14,15] This first episode will serve as a context for exploring gesture research literature in more detail in Section IV.

A. Trajectory episode

In this section, I draw examples of the use of gestures from a conversation among three students (pseudonyms Mike, Robin, and Jenny) working together with a teaching assistant on their mechanics homework. In the minute-and-a-half that is analyzed in detail, the students are trying to draw velocity vectors at various points along the trajectory of an object in projectile motion. In particular, the students are discussing whether the velocity of the object is zero at the top of its trajectory. In this case, the velocity of the object at the top of its trajectory is horizontal.

In the first thirty-five seconds of the 1.5-minute episode, the students are unsure about the velocity of the object at the top. In the next thirty seconds, the TA guides the students to think about the components of the velocity vectors for the object. At the end of the episode, the group agrees that at the top the object has horizontal velocity, but not vertical. A transcript of the episode follows. (TA, J, M, and R are the teaching assistant, Jenny, Mike, and Robin respectively.)

*Video 1*

TA: And you said the one at the top should have zero length? Is that –

J: We weren't going to draw an arrow.

M: We're not – we're not sure what to do.

J: Since the velocity's zero at that one point

TA: Is it not moving at that point?

J: Right.

TA: Why?

J: It kinda comes to a s- – well,

M: Changing direction?

J: Does it come to a stop at that point? If it stopped

R: There's one instant when it ha-

J: So it just-

TA: Well, let's try this. Let's just try – I mean – to break up the two different directions. I think you've done this in class where you draw, um, vectors in different directions you break them up into components. Have you done that in class? Where you have like a vertical and a horizontal

R: Kinda.

J: I don't know what we've done in class.

M: I'm trying to pick up where you're going with this.

J: Yeah.

TA: Okay, where I'm going. If you just think about it moving left to right, does it ever stop when it's moving left to right?

R: No.

TA: Why?

Scherr

J: Wouldn't it just fall straight down then if it was like –Wmp! Psh. [gestures]

M: [Laughs] I don't know.

J: If it stopped and then it – I'm trying to think.

M: It still has I guess horizontal motion, but not...

TA: Not vertical.

J: Vertical.

M: It's not increasing its...height.

J: Right.

Jenny, Mike, and the TA all gesture during the episode. Four gestures are presented here as examples of the type of spontaneous hand motions that are identified as gestures. In the first line, while saying "the one at the top should have zero length," the TA holds his arm out straight toward the picture of the trajectory that is on the board above the students and holds his index finger and thumb close together, as though to indicate something small (Figure 1a). When he suggests thinking about "two different directions," he holds his two hands flat and at right angles to one another (Figure 1b). For "vectors," he slides the tip of his right index finger along and then beyond his pointed left index finger (Figure 1c). Jenny shrugs at least twice (a gesture analyzed extensively in Ref. 1), and at the end of the episode Mike shoots his outstretched arm straight upwards while saying "It's not increasing its…height" (Figure 1d). Other gestures appear in the episode as well.

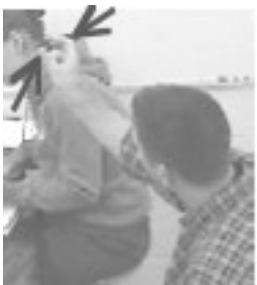

FIG. 1a. "zero length"

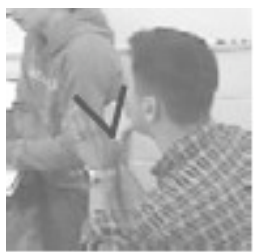

FIG. 1b. "two different directions"

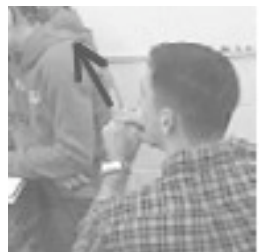

FIG. 1c. "vectors"

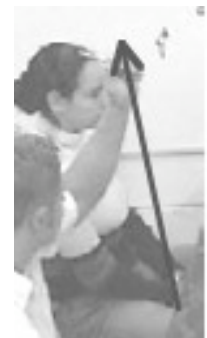

FIG. 1d. "It's not increasing its …height"

B. The "half-parabola" gesture and its apparent source

One gesture by Jenny, the only gesture marked in the transcript, is particularly significant to the conversation. As she says "Wouldn't it just fall straight down then if it was like –Wmp! Psh," her left hand moves up in front of her body in a half-parabola ("if it was like"), stops at the vertex ("Wmp!"), and finally drops straight down to the table ("Psh"). Figure 1e is a still frame illustrating the gesture.

One measure of the gesture's significance is that her statement is unintelligible without it. Another measure of the gesture's significance is that it is an intuitively compelling expression of Jenny's thinking about the motion, and the participants treat it as such; at Jenny's gesture, the TA stops talking, and within a few seconds the group reaches the correct conclusion about the velocity at the top of the trajectory. Finally, the gesture is eloquent; it's hard to imagine words that would complete Jenny's sentence with anything like the clarity and brevity that the gesture provides.

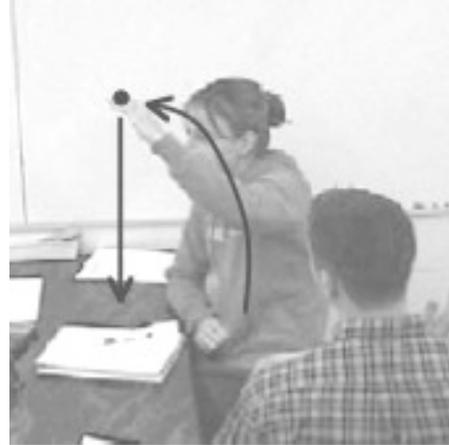

FIG. 1e. The half-parabola gesture.

The half-parabola gesture appears at first to function, conversationally, as an answer to the homework question – or at least a strong, common-sense refutation of an answer that was competing with the correct answer. Also, appearing as it does immediately after the TA intervention about components, the gesture seems to be a result of that intervention. It is tempting to conclude that the TA intervention was a helpful one; it seems to have elicited the half-parabola gesture, which gave the students a common-sense basis for the correct response.

However, closer examination of the videotape reveals that the cited occurrence of the half-parabola gesture was not, in fact, its first occurrence. In the first half of the episode, about forty seconds before the event described above, Jenny's hand makes the half-parabola gesture twice as she says quietly: "If it stopped… so if it stopped…" Jenny's incomplete sentences function as musing to herself (since no one responds to her); the TA overlaps her musing to begin his intervention about components.

Since the gesture first occurred before the TA intervention, it was surely not a result of that intervention – and perhaps the intervention was not as helpful as one might have thought, at least to Jenny. On the contrary, for Jenny, the intervention appears to have been an interruption in her thinking. Closer examination of the video reveals that Jenny appears to be waiting for the TA to finish so that she can speak; just after the initial gestures, she draws a breath as though to begin talking, but then looks at the TA and closes her mouth tightly.

The above example illustrates some of the more obvious benefits of attending to student gestures. As researchers, we can use the initial appearances of the half-parabola gesture to distinguish ideas that Jenny generated herself from ideas that were precipitated by the TA intervention. In addition, had the TA in this episode been alert to Jenny's gesture, he might not have drawn her away from her own thoughts with an interjection about components. Instead, he might have inquired about the initial occurrences of the half-parabola gesture, perhaps by imitating and questioning it: "What do you mean by this?"

## IV. GESTURE RESEARCH REGARDING IDEA CONSTRUCTION

The half-parabola gesture is particularly interesting in that it appears to represent an idea that Jenny constructs herself during the course of the conversation. As researchers (and as instructors), we want to be able to recognize moments when students are engaged in the construction of ideas, and distinguish those from moments when students are recounting ideas they learned earlier. This distinction is not always an easy one to make; for example, students sometimes make statements that sound fresh and convincing only to reveal later that they were parroting poorly-understood textbook material. Gestures offer one source of evidence of students' engagement in constructive thinking. This section reviews existing gesture research regarding idea construction, using the half-parabola gesture described above as a touchstone example.

### A. Indicating "pre-articulate" ideas

One of the key features of the half-parabola gesture is that it appears well before Jenny's idea is articulated verbally (if her idea is ever articulated verbally at all). Gestures can convey scientific ideas that are not yet articulate: for example, a student may enact individual forces in a pulley system correctly at a point when his verbal and diagrammatic presentation is still scientifically inaccurate.[16] The delay between a gesture and the corresponding scientifically-correct words may initially be considerable (sometimes weeks); as learners gain experience, the gap closes until the gesture and corresponding talk occur simultaneously.[17] Such results strengthen our sense that because Jenny gestures before (or instead of) speaking, the gesture illustrates an emerging idea.

### B. Distinguishing "explanation" from "description"

Some gesture research claims to distinguish students' "descriptions" of a memorized or previously thought-out model from "explanations" constructed in the moment by observing the timing of gestures as well as the *gestural viewpoint*, that is, whether the gesturer is inside or outside the gesture space.[18] In such an analysis, Jenny's gesture would identify her expression as more "explanatory" than "descriptive," first because the gesture is interwoven with speech and second because the gesture is large enough to include Jenny herself in its expanse. Unfortunately, the cited study is weakened by the absence of an independent measure distinguishing "description" from "explanation."

### C. Displaying the novelty of ideas

Other gesture research demonstrates methods by which we may at least identify ideas that are being treated as new for purposes of communication, whether or not they are new to the speaker. When speakers describe a previous action to an addressee who has done the same actions (in this case, played with distinctive toys), these "common-ground" gestures are significantly less precise, complex, or informative than when the listener does not share common experiences.[19] Similarly, in a series of gestures about the same actions by the same speaker (*e.g.*, increasingly detailed descriptions of a certain toy), speakers emphasize new information in each gesture by making that aspect of the gesture larger or clearer. When information becomes "given" (rather than new), a gesture for the same information becomes smaller or less precise. The researchers conclude that the immediate communicative function of the gesture plays a major role in determining the gesture's physical form. For our purposes, it is equally of interest that the form of the gestures indicates the extent to which the speaker understands the communicated idea to be new to her listeners.

The above studies do not document shifts in gestural viewpoint, nor do they directly address how gestures may identify ideas that are new to the speaker (*i.e.*, "explanations" rather than "descriptions"). It is plausible to assume that ideas that are new to the speaker would be treated as new to the listener as well, and would be evidenced by relatively large, precise, complex, and informative gestures. However, the converse is not true: ideas that are familiar to the speaker are not necessarily accompanied by "common ground" gestures. The form of the gesture is influenced by the speaker's perception of whether the listener shares common experiences. In a physics classroom, it is reasonable to guess that students will assume instructors have already heard any physics idea that they are reciting from memory. Thus, it is probably safe to assume that "new"-type gestures that students make to instructors indicate ideas that are new to the students.

### D. Facilitating idea construction

Gesture facilitates thought and speech, as well as illustrating them. Physicists and chemists in particular have been observed to arrive at their understanding of specific topics in part through gesturing.[20] Numerous studies affirm that speakers "offload" thinking into gesture as into a sketch, freeing up cognitive effort for other tasks.[21] Gesturing in order to explain a math problem, for example, improves participants' ability to simultaneously remember a list of words or letters.[22] Physics problem-solving can be quite difficult, often requiring students track complicated chains of inference by holding complex structures of ideas in their minds and manipulating or reasoning about them. Jenny's externalized "air drawing" of the hypothesized trajectory may enable her to think *about* the trajectory rather than merely *of* it.

The mechanism by which gesture facilitates idea construction is the subject of current research. Some results suggest that gesturing facilitates word retrieval.[23] Indeed, gesture originates in the same regions of the brain as language and activates the same neuronal assemblies as the spoken language that goes with it, making it plausible for gesture to assist neurologically with the process of speaking and with language emergence more generally.[24] However, other results suggest that gesture is involved not only in retrieval of ideas but in the conceptual planning of speech – in particular, that gesture helps speakers to organize spatial information into units that are more easily verbalized.[25] This evidence indicates that gesture plays an integral role not only in speaking, but also in thinking.

Another possible mechanism by which gestures may facilitate idea construction is kinesthetic. The fact that Jenny physically moves her hand in order to illustrate her thinking may give her sensorimotor information that influences the development of her idea.[26] If she acts out the conjecture that the velocity of the projectile is zero at the top, the consequence – that the projectile would drop straight down instead of continuing on the given trajectory – might follow without any semantic analysis. Jenny might simply *feel* what the object would do. Young children in particular are systematically observed to perform gestures that are close to being *enactments,* having many features in common with physical actions performed on the external environment;[27] it is easy to imagine them making inferences about those actions based on kinesthetic feedback from gesture. Another anecdote illustrates a similar possibility in a different way: A student was trying to decide whether dropping her keys from a position above her foot would result in the keys falling onto her foot if she were in a chair rolling with constant velocity. She initially thought the keys would fall in front of her foot (due their initial forward velocity), but when she sat in the rolling chair and held her hand out over her foot, she immediately said the keys would hit her toe.[28] She had no need to do the experiment. The sensation of holding the keys in the moving chair gave her information that changed her prediction. Theoretical frameworks including distributed cognition (in which the unit of analysis for cognition includes not only individual brains, but also bodies, material structures, and social contexts)[29] and embodied cognition (in which our bodily experience of the world both enables and constrains conceptual understanding)[30] support the idea that gestures provide information that can influence thinking.

## V. THIRD LAW EPISODE: GESTURES AS EVIDENCE OF THE NOVELTY AND CONSTRUCTION OF STUDENT IDEAS

The episode below illustrates changes in the gestures that students produce as they make a series of statements about a collision. The changes in their gestures seem to indicate changes in the status of the ideas to which the gestures refer, from perhaps being newly constructed at the outset to being familiar territory later on.

### A. Third law episode

In this episode, four students (three of whom speak) are working on a tutorial on Newton's third law. The tutorial begins by stating the third law and admitting that in some cases it seems not to make sense (an admission that is well supported by research into student understanding of Newtonian mechanics[31]). Students are asked to consider a heavy truck ramming into a parked, unoccupied car, and are asked, "According to common sense, which force (if either) is larger during the collision: the force exerted by the truck on the car, or the force exerted by the car on the truck?"[32]

In the video below (Video 2), "Kendra" (K) expresses a common concern about whether the third law applies to the situation described; "Alan" (A) and "Jasmin" (J) try to reconcile the third law with the common-sense intuition that the force by the truck would be larger. A few minutes later (Video 3), the students explain their initial idea to a teaching assistant (TA). After the teaching assistant leaves, the students continue to wrestle with the apparent contradiction (Video 4). Below, transcript of each video is accompanied by description of some the gestures that appear in each video. The subsequent section discusses the evidence that the gestures provide regarding the students' construction of physics ideas.

*Video 2*

K: I could never understand that, but... does this go against the law then, or is it that they are equal but we just think it's the truck? You understand what my question...

A: We think it's the truck because the truck doesn't move backwards, I think. Right? Cause if there's equal and opposite forces, the truck... we would, if we actually saw it, we'd think the truck would hit the car and go backwards because of the force, but since...

J: Maybe they do exert the same force, but the truck doesn't move.

A: The truck doesn't move cause I think it's got the momentum going, and... you know.

K: So they, they are doing the same force, it just doesn't... it's just not common sense.

A: Yeah, it just doesn't register because we see the truck (K: It looks as being bigger) still moving forward, right, right.

In Video 2, Alan gestures prolifically. On "*We think* it's like the truck," Alan points to his temple (Figure 2a). On "the truck doesn't *move backwards*," he pulls his two cupped hands towards his body (Figure 2b). For "equal and opposite forces," Alan's palms face each other, tracing a curve downward and outward from a central location in front of his body to land vertically on the table (Figure 2c). "The truck" is accompanied by vertical zigzags of his open hands (Figure 2d). "If we actually *saw it*" is indicated by a pointing finger moving out from Alan's eye (Figure 2e). For "we'd think the truck would *hit the car* and *go backwards*," Alan's left palm smacks his right palm (Figure 2f), then "rebounds" some distance to the left (Figure 2g). Jasmin, shortly afterward, says "Maybe they do exert the *same force,*" and pinches each index finger and thumb together, holding the left- and right-hand pinches together in a point (Figure 2h).

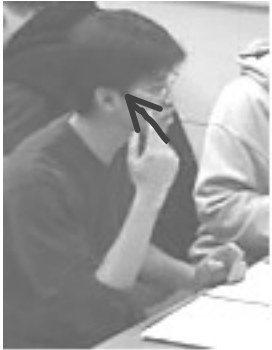
FIG. 2a. "we think"

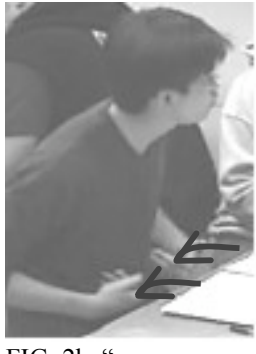
FIG. 2b. "move backwards"

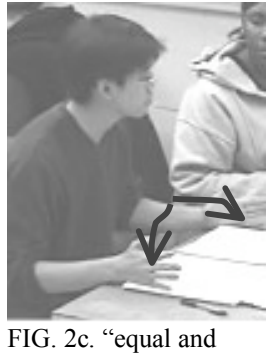
FIG. 2c. "equal and opposite forces"

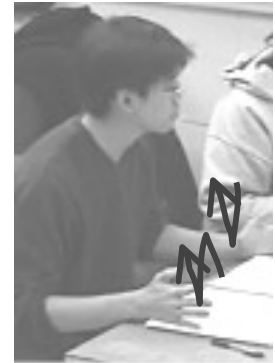
FIG. 2d. "the truck"

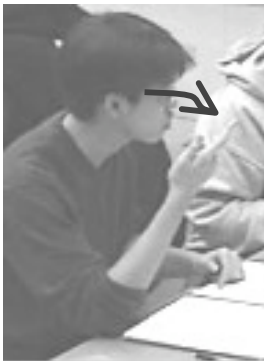
FIG. 2e. "saw it"

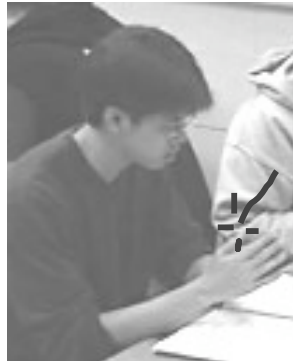
FIG. 2f. "hit the car"

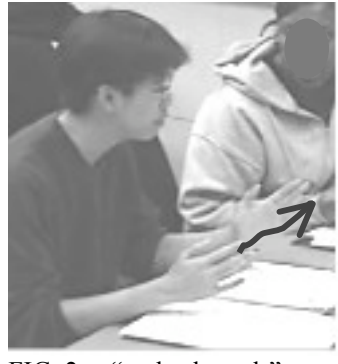
FIG. 2g. "go backwards"

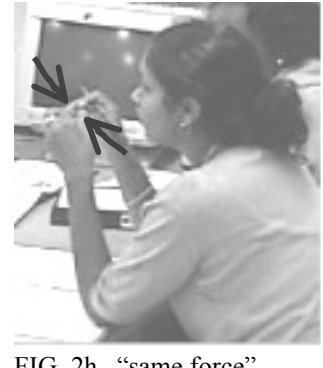
FIG. 2h. "same force"

*Video 3*

TA: What did you say initially, what did you guys think would happen?

K: That the truck would have...

A: The truck, OK.

K: The truck would have a greater force. I mean, that's our common sense (A: common sense would say that), but it just goes against the Newton's law.

TA: OK, well, we're gonna... today we're gonna try and look at that.

J: No, we were saying that maybe they do exert the same force, but the truck

A: has just more, I dunno, speed, momentum, whatever, so it doesn't move backwards.

In Video 3, Jasmin and Alan both make gestures worth noting. When Jasmin says, "we were saying that maybe they do exert the *same force*," she pinches the index finger and thumb of one hand together (Figure 3a). When Alan says "has just more, I dunno, speed, momentum, whatever, so it doesn't *move backwards*," he waves both hands loosely

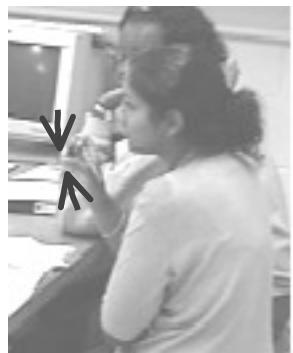
FIG. 3a. "same force"

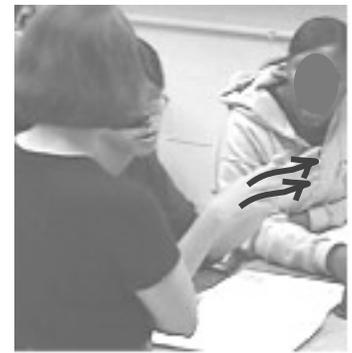
FIG 3b. "move backwards"

to his left with palms facing each other (Figure 3b).

*Video 4*

J: I mean, it doesn't seem right. I get it, I totally understand what you're saying.

A: Yeah, yeah yeah.

K: Would you just say that there must be some type of plausible explanation about it?

J: No, I think that they are exerting the same force on each other, but it just looks like, when we look at it...

A: When you look at it, you see the other car running over the smaller car.

K: Maybe if you were blind, you wouldn't... it's just cause your eyes and common sense (A: right, rightright) plays a role in it, that's why.

A: But then again, like sometimes you see, like...

J: I wonder if that truck felt the same... the person in the truck feels the same thing?

A: Feels the same force.

In Video 4, notable gestures happen toward the end of the video. Jasmin first rocks her upper body forward, supporting herself on the edge of the table (Figure 4a), and then pushes backward away from the table (Figure 4b), as she says "I wonder if that truck felt the same... the person in the truck feels the same thing?" Alan, in response, states "Feels the same force" as the fingers of his upward-facing left hand close loosely onto the palm (Figure 4c).

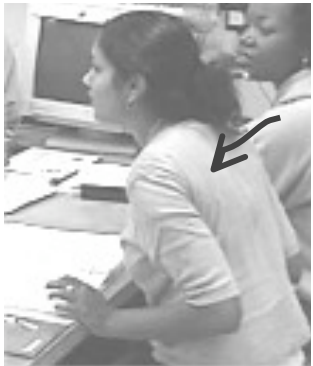 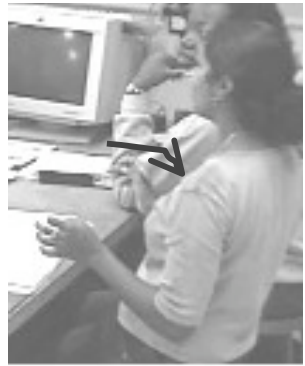 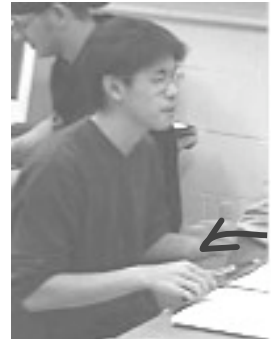

FIG. 4a. "I wonder if that truck felt the same"

FIG. 4b. "the person in the truck feels the same thing"

FIG. 4c. "feels the same force"

B. Significance of gestures in the third law episode

In the third law episode, none of the gestures are substitutes for speech; we can understand the content of the students' ideas without knowing their hand motions. However, the gestures in this episode are evidence of the status of the students' ideas. In particular, they indicate that certain ideas are relatively new to the students at the outset and are experienced as more familiar to the students later on. The gestures may also be facilitating students' construction of ideas, as well as indicating their relative novelty.

*Gestural shifts indicating decreasing novelty of ideas*

In Video 2, Alan's gestures are prolific, detailed, large, and strongly animated relative to his gestures in later videos, suggesting that he is explaining ideas that he considers to be new, either to him or to his listeners (Refs. 18 and 19). For example, in Video 2, he illustrates "move backwards" with cupped hands moving toward his body (Figure 2b), suggesting that the truck in this imagined instance is moving backwards into his torso. In Video 3, Alan illustrates "move backwards" by waving both hands loosely to his left (Figure 3b); the gesture is smaller, less well defined, and no longer includes his body in its space. Similarly, Jasmin's gesture for "same force" begins as a two-handed gesture (Figure 2h), but is one-handed when later performed for the teaching assistant (Figure 3a). These changes in a gesture's form are consistent with the changes observed in Ref. 19 for gestures representing ideas of decreasing novelty. The change in gestural viewpoint is consistent with changes observed in Ref. 18 regarding shifts from "explaining-in-the-moment" to "describing" previously thought-out models.

Alan's gesture in Video 4, accompanying "feels the same force" (Figure 4c), may be a reduced version of the gesture accompanying "move backwards" in Video 2 (Fig 3b); it uses one loose hand instead of two strong ones, but the handshape and motion toward the body is similar. The fact that it's the left hand that makes the gesture recalls the left-hand emphasis of the "go backwards" gesture from Video 2 (Figure 2g) as well as the "move backwards" that appears in Video 3 (Figure 3b). In those earlier references, the hypothetical backwards motion of the truck was either into Alan's body, or to the left; in the latest gesture (Figure 4c), the left hand flicks towards Alan's body, possibly referencing both "backwards" directions of the previous gestures in a single new gesture. The fact that the latest gesture accompanies the words "feels the same force" may serve to communicate that it's the *truck* that's feeling the same force – the same truck that "doesn't move backwards" in earlier references. Gestures, being free from the constraints of syntax imposed on language, are free to participate in imagistic blends such as this one.[33]

*Gestures potentially facilitating idea construction*

The particular ideas these students are exploring in these episodes are notoriously difficult to learn.[31] In addition, the conceptual pathway laid out by the worksheet that guides their activity is complicated, involving exploration of various hypothetical possibilities that are in some cases quite counterintuitive. To the extent that gesturing can free up cognitive effort for the task at hand, Alan's and Jasmin's numerous gestures may help them to think more clearly.[21] In particular, the gestures may help to activate the scientific language associated with the scenario by organizing the wealth of spatial and movement information suggested by it.[23-25] Jasmin's whole-body gesture, rocking forward and then pushing herself away from the desk, is particularly evocative. Her enactment of what the person in the truck might feel potentially gives her kinesthetic information with which to evaluate the plausibility of their conclusions about the collision.[26,29,30]

# VI. OTHER GESTURE RESEARCH OF INTEREST: GESTURE-SPEECH MISMATCHES

A rich area of gesture research that is not addressed in either of the above two episodes is *gesture-speech mismatches,* in which speakers make gestures that are inconsistent with their speech.[34] For example, children performing Piaget conservation tasks use gestures in their explanations, and sometimes those gestures contradict their verbal descriptions (*e.g.,* indicating "wider" when the child said "taller").[35] The researchers in the cited study classified the students as to whether they were conserving or nonconserving (*i.e.,* whether they correctly identified quantity invariance under displacement transformation) and also as to whether their gestures were "concordant" or "nonconcordant" (*i.e.,* whether the information conveyed in gesture matched that conveyed in speech). In a subsequent teaching task, nonconserving nonconcordant children learned the most. The suggested explanation is that gesture-speech mismatches indicate when speakers are "of two minds," and have both conservation and nonconservation cognitively available for instructional reinforcement. Another study observed that gesture-speech mismatches cluster at strategic choice points in solving the Tower of Hanoi puzzle, suggesting that cognitive exploration of more than one solution path is taking place.[36] Gesture-speech mismatches appear to be an even better predictor of subsequent learning than other forms of inconsistency, such as multiple inconsistent verbal statements, suggesting that the inconsistency across modalities is a particularly significant index of transitional knowledge.[37] Although I have not presented examples of gesture-speech mismatches in this article, such occurrences would be of great interest in understanding physics students' thinking as well as their readiness to learn.

# VII. CONCLUSION

Research on gesture analysis includes much to interest physics education researchers. Some of the research is potentially of immediate use in diagnosing student thinking, assisting physics education researchers in identifying not only the content of student ideas, but also their source, their novelty to the speaker, and whether the speaker is actively engaged in constructing the ideas. Other research has pedagogical and theoretical implications for our understanding of how learning occurs, offering, for example, an account of how speakers use gesture to reduce cognitive load, or how gesture may assist with organizing information. Physics education is a rich field for exploring these issues further, and physics education researchers may both benefit from and contribute to continuing investigations of the significance of gesture in thinking and learning.

# ACKNOWLEDGEMENTS

The work described in this paper has been a collaborative effort by many members of the Physics Education Research Group and other colleagues. I am especially grateful for the intellectual contributions of David Hammer. This research is supported in part by the National Science Foundation Grant No. REC 0440113.